\documentclass[conference]{IEEEtran}
\IEEEoverridecommandlockouts
\usepackage{cite}
\usepackage{amsmath,amssymb,amsfonts}
\usepackage{algorithmic}
\usepackage{graphicx}
\usepackage{textcomp}
\usepackage{xcolor}
\usepackage{lipsum}
\def\BibTeX{{\rm B\kern-.05em{\sc i\kern-.025em b}\kern-.08em
    T\kern-.1667em\lower.7ex\hbox{E}\kern-.125emX}}
\begin{document}

\title{AI-Guided Exploration of Large-Scale Codebases
}

\author{\IEEEauthorblockN{Yoseph Berhanu Alebachew}
\IEEEauthorblockA{\textit{Department of Computer Science} \\
\textit{Virginia Tech}\\
yoseph@vt.edu}
}

\maketitle



\section{Introduction}
Modern software systems are large, complex, and continuously evolving. As systems grow in size and interconnectivity, developers face increasing challenges in understanding unfamiliar codebases—whether during onboarding, debugging, refactoring, or feature development. Prior research estimates that professional developers spend up to 70\% of their time on program comprehension tasks \cite{xia2018measuring}. 
This problem is especially pronounced in real-world systems involving layered abstractions, distributed modules, and sparse or inconsistent documentation.

A wide range of strategies have been studied to support program comprehension, including static analysis, software visualization, reverse engineering, and summarization. However, existing tools remain inadequate in practice. Many rely on static views that fail to support the adaptive, dynamic, interactive strategies developers use in real-world settings. 
These tools often lack interactivity, do not scale with project size, and force developers to context-switch between tools to retrieve structural, historical, or semantic information—leading to increased cognitive load.

Recent advances in large language models (LLMs) offer new opportunities for augmenting comprehension workflows. LLMs can generate visualization, natural-language summaries, suggest exploration paths, or answer questions about unfamiliar code. Yet their use in program comprehension remains limited by concerns over accuracy, lack of grounding, and poor integration with interactive visual interfaces. In particular, their non-deterministic output and limited access to fine-grained program structure make them difficult to trust or scale.

This work contributes to ongoing research on interactive, intelligent tools for code understanding by exploring how LLMs can be integrated with deterministic reverse engineering techniques to support adaptive, multi-level, and context-aware comprehension. Our focus is not only on generating diagrams, but on enabling a more fluid and guided interaction between developer and system—where structural representations, semantic insights, and contextual information are combined to support effective exploration of large-scale codebases.

\section{Relation to Prior Work}
Our work builds upon a rich body of research in software visualization and program comprehension. Traditional tools such as SHriMP~\cite{storey2001shrimp}, CodeCity~\cite{wettel2008city}, and CodeCompass~\cite{vass2015codecompass} provide static visualizations of software structure using hierarchical or metaphor-based models. These tools help developers form mental models of code but often fall short in supporting dynamic comprehension strategies, collaborative contexts, or integration of historical and contextual information.

Pacione et al.~\cite{pacione2004novel} proposed a multi-faceted visualization model supporting different levels of abstraction and perspectives (structure, behavior, data). While influential, this model lacks real-time interaction and adaptability to user exploration patterns. Similarly, tools like Azurite~\cite{nguyen2013azurite}  visualize code evolution over time, yet they remain disconnected from comprehension strategy modeling or semantic context.

Interactive interfaces such as CodeBubbles~\cite{bragdon2010codebubbles} improved support for developer cognition by offering flexible and visual engagement with code segments. However, these tools do not integrate reverse engineering, natural language guidance, or collaborative features. By incorporating collaborative features, our work also relates to collaborative comprehension tools like LiveShare\footnote{https://visualstudio.microsoft.com/services/live-share/}  and GitLive\footnote{https://git.live/}, which support real-time team navigation but do not offer architectural visualization or semantic understanding capabilities.

Our approach advances the literature by integrating an LLM agent into the visualization loop. We propose moving beyond static diagrams to support adaptive, interactive, and intent-aware navigation. The LLM acts as a reasoning and mediation layer that can respond to structured interactions (clicks, filters, annotations) and suggest exploration paths or summarize historical changes. This makes our system one of the first to unify static code structure, dynamic visualization, semantic context extraction, and collaborative UI guidance under a single runtime framework.

In summary, our work bridges the gap between static visualization models and emerging conversational agents by introducing an LLM-guided, multimodal interface for program comprehension. It synthesizes ideas from reverse engineering, information foraging~\cite{pirolli1999information}, cognitive load theory, 
and experience design
into a coherent tool for exploring large-scale systems.

\section{AI-Supported Visual Exploration}

Our approach combines deterministic reverse engineering with LLM-guided, intent-aware interaction to support scalable program comprehension. The system architecture consists of four core components:

\begin{itemize}
    \item \textbf{Code-to-UML Reverse Engineering:} Source code is parsed using abstract syntax trees~\cite{sun2023ast} and structural reduction techniques 
    to generate UML diagrams. 
    These diagrams support multi-level abstraction and modular decomposition, enabling both top-down and bottom-up comprehension.

    \item \textbf{Interactive Visualization:} The front-end renders dynamic visualizations with interactive features such as zooming, panning, and drill-down (e.g., clicking a high-level module to explore its internals). Additional layers include overlays (e.g., change frequency heatmaps), filters 
    and historical comparisons across versions.

    \item \textbf{LLM-Guided Interface Planner:} The LLM interprets user queries and interactions to recommend guided exploration paths, provide contextual summaries, and dynamically update the interface via a structured JSON-based GUI specification. For instance, a user might request ``show all modules modified in the last release'' or inspect the history of a specific component. The planner can incorporate prior exploration traces—both from the current user and shared team workflows—to improve its guidance over time.

    \item \textbf{Context and Collaboration Layer:} Visualizations are enriched with contextual information from version control artifacts. The system also supports collaborative features, including shared views, real-time annotations, and embedded documentation, allowing distributed teams to align understanding and maintain a shared mental model of the codebase.
\end{itemize}

Together, these components form a closed interaction loop: static code structure is visualized; the user explores or poses a query; the LLM interprets intent and context to refine the view; and the updated visualization informs the next step in comprehension. This iterative cycle transforms code exploration into an adaptive, multimodal process that integrates structural, semantic, and social signals.

We have developed a functional prototype demonstrating the feasibility of this approach, including support for multi-level diagrams and contextual exploration. Our current implementation supports Java programs, but the approach can easily be extended to support other languages. 

\section{Future Work}

Our prototype validates the feasibility of integrating LLMs with reverse engineering to support program comprehension. As a next step, we will conduct a user study to evaluate how the system impacts comprehension accuracy, task completion time, and perceived cognitive load, compared to baseline tools. Participant feedback will guide refinements to the interaction flow, the interface usability, and the level of trust in LLM-generated suggestions.

Following this evaluation, we aim to extend the system to handle larger and more complex codebases, integrate runtime behavior analysis, and support real-time collaborative exploration. These improvements will help broaden the applicability of AI-assisted comprehension tools in professional development workflows.

Finally, we plan to investigate Graphical User Interface (GUI)--based interaction as a primary modality for LLM integration—moving beyond traditional chat interfaces—to support more intuitive, context-sensitive exploration aligned with developer intent in tasks other than comprehension (e.g., debugging, code review). Several open questions remain, such as how to ensure the accuracy and trustworthiness of LLM-generated guidance, how to manage long-term interaction context effectively, and how to scale the system to massive polyglot codebases. Addressing these challenges will be central to advancing the applicability and reliability of AI-assisted comprehension tools in real-world settings.

\section{Conclusion}
We propose an AI-guided approach to large-scale program comprehension that bridges traditional reverse engineering with modern LLM-based interaction. By combining structural visualization with conversational guidance, we aim to reduce developer cognitive load and enable a more intuitive, strategic exploration of software systems. Our prototype lays the groundwork for future tools that treat software understanding as an adaptive, collaborative process.

This hybrid design opens new possibilities for augmenting developer workflows through intelligent assistance that is grounded in both code structure and human intent. As the field continues to explore AI-driven tooling for software engineering, we envision this approach contributing to a new generation of environments that are explainable, user-driven, and closely aligned with the way developers think and work.

\bibliographystyle{IEEEtran}
\bibliography{references}
\end{document}